\documentclass[twocolumn,english,prl]{revtex4}
\usepackage{amsmath}
\usepackage{graphicx}
\usepackage{amssymb}
\usepackage{bm}
\usepackage{babel}
\begin{document}

\title{Universal destabilization and slowing of spin transfer functions
by a bath of spins}

\author{Daniel Burgarth and Sougato Bose}

\affiliation{Department of Physics \& Astronomy, University College London, Gower
St., London WC1E 6BT, UK}

\begin{abstract}
We investigate the effect of a spin bath on the spin transfer functions
of a permanently coupled spin system. When each spin is coupled to
a seperate environment, the effect on the transfer functions in the
first excitation sector is amazingly simple: the group velocity is
slowed down by a factor of two, and the fidelity is destabilized by
a modulation of $\left|\cos Gt\right|,$ where $G$ is the mean square
coupling to the environment.
\end{abstract}

\maketitle

\paragraph*{Introduction:---}

Recently suggested protocols \cite{Sougato,C1,NJP} give a new perspective
to the physics of strongly coupled spin systems. They demonstrate
that the coherent transfer of spin flips can be used to transfer unknown
quantum states and entanglement, a task of paramount importance in
any quantum information application \cite{Nielsen}. Generally, the
relevant quantities determining the performance of the mentioned protocols
are the time dependent transition amplitudes of local spin flips in
a ferromagnetic ground state. We will refer to these amplitudes as
{}``spin transfer functions''. The same functions also occur in the charge and energy transfer dynamics in molecular systems \cite{EXCITON} and in continuous time random walks \cite{OLI} to which our results equally apply.

It is both important and interesting to ask how these transfer
functions change if the intended couplings between the spins are
accompanied by unwanted couplings to environmental spins which do
not take part in the transport. It is well known from the theory
of open quantum systems \cite{key-36} that this can lead to
dissipation and decoherence, which also means that quantum
information is lost. Here we consider a model where the system is
coupled to a spin environment through an exchange interaction
because the same type of coupling is also responsible for the
transport of the information through the system. Moreover, this
coupling offers the unique opportunity of an analytic solution of
our problem without {\em any} approximations regrading the
strength of system-environment coupling (in most treatments of the
effect of an environment on the evolution of a quantum system, the
system-environment coupling is assumed to be weak) and allows us
to include inhomogeneous interactions of the bath spins with the
system. For such coupling, decoherence is possible for mixed
(thermal) initial bath states \cite{key-33}. However if the system
and bath are both initially cooled to their ground states, is
there still a non-trivial effect of the environment on the spin
transfer functions? In this paper we find that there are two
important effects: the spin transfer functions are slowed and a
destabilized due to the environment. This has both positive and
negative implications for the use of strongly coupled spin systems
as quantum communication channels.

\paragraph*{Model:---}

\begin{figure}[htbp]
\begin{center}\includegraphics[%
  width=1\columnwidth]{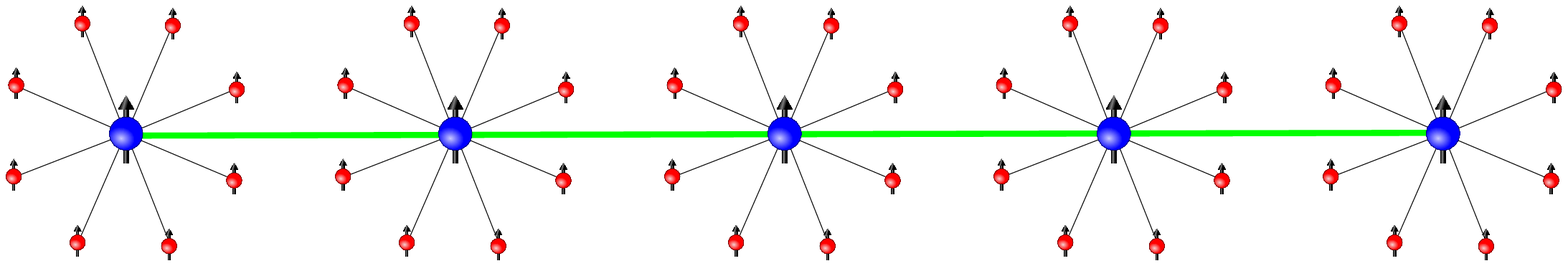}\end{center}

\caption{\label{fig:spinchain}A spin chain of length $N=5$ coupled to independent
baths of spins. }
\end{figure}

We choose to start with a specific spin system, i.e. an open spin
chain of arbitrary length $N,$ with a Hamiltonian given
by\begin{equation}
H_{S}=-\frac{1}{2}\sum_{\ell=1}^{N-1}J_{\ell}\left(X_{\ell}X_{\ell+1}+Y_{\ell}Y_{\ell+1}\right),\end{equation}
where $J_{\ell}$ are some arbitrary couplings and $X_{\ell}$ and
$Y_{\ell}$ are the Pauli-X and Y matrices for the $\ell$th spin.
Towards the end of the paper we will however show that our results
hold for any system where the number of excitations is conserved
during dynamical evolution. In addition to the chain Hamiltonian,
each spin $\ell$ of the chain interacts with an independent
bath of $M_{\ell}$ environmental spins (see Fig \ref{fig:spinchain}%
) via an inhomogeneous Hamiltonian,\begin{equation}
H_{I}^{(\ell)}=-\frac{1}{2}\sum_{k=1}^{M_{\ell}}g_{k}^{(\ell)}\left(X_{\ell}X_{k}^{(\ell)}+Y_{\ell}Y_{k}^{(\ell)}\right).\end{equation}

\begin{figure}[htbp]
\begin{center}\includegraphics[%
  width=1\columnwidth]{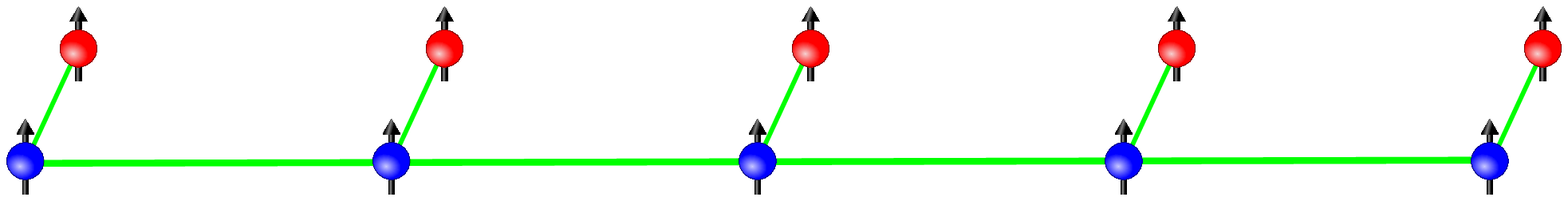}\end{center}

\caption{\label{fig:spinchain_equiv}In the first excitation sector, the 
system can be mapped into an effective spin model where the bath spins
are replaced by a single effective spin, as indicated here for $N=5.$}
\end{figure}

In the above expression, the Pauli matrices $X_{\ell}$ and $Y_{\ell}$
act on the $\ell$th spin of the chain, whereas $X_{k}^{(\ell)}$
and $Y_{k}^{(\ell)}$ act on the $k$th environmental spin attached
to the $\ell$th spin of the chain. We denote the total interaction
Hamiltonian by\begin{equation}
H_{I}\equiv\sum_{\ell=1}^{N}H_{I}^{(\ell)}.\end{equation}
 The total Hamiltonian is given by $H=H_{S}+H_{I},$ where it is important
to note that $\left[H_{S},H_{I}\right]\neq0.$ The ground state of
the system is given by the fully polarized state $|0,0\rangle,$ with
all chain and bath spins aligned along the z-axis. The above Hamiltonian
describes an extremely complex and disordered system with a Hilbert
space of dimension $2^{N+NM}.$ In the context of state transfer however,
only the dynamics of the first excitation sector is relevant. We proceed
by mapping this sector to a much simpler system \cite{Alexandra,Alexandra2}.
For $\ell=1,2,\ldots,N$ we define the states\begin{equation}
|\ell,0\rangle\equiv\sigma_{\ell}^{+}|0,0\rangle\end{equation}
and\begin{equation}
|0,\ell\rangle\equiv\frac{1}{G_{\ell}}\sum_{k=1}^{M_{\ell}}g_{k}^{(\ell)}\sigma_{k}^{+(\ell)}|0,0\rangle\end{equation}
with \begin{equation}
G_{\ell}=\sqrt{\sum_{k=1}^{M_{\ell}}\left(g_{k}^{(\ell)}\right)^{2}}.\label{eq:effect}\end{equation}
 It is easily verified that\begin{eqnarray}
H_{S}|\ell,0\rangle & = & -J(1-\delta_{\ell1})|\ell-1,0\rangle-J(1-\delta_{\ell N})|\ell+1,0\rangle\nonumber \\
H_{S}|0,\ell\rangle & = & 0,\label{eq:hc_action}\end{eqnarray}
and\begin{eqnarray}
H_{I}|\ell,0\rangle & = & -G_{\ell}|0,\ell\rangle\label{eq:hb_action}\\
H_{I}|0,\ell\rangle & = & -G_{\ell}|\ell,0\rangle.\label{eq:hb_action2}\end{eqnarray}
 Hence these states define a $2N-$dimensional subspace that is invariant
under the action of $H.$ This subspace is equivalent to the first
excitation sector of a system of $2N$ spin $1/2$ particles, coupled
as it is shown in Fig \ref{fig:spinchain_equiv}. %
Our main assumption is that the bath couplings are \emph{in effect}
the same, i.e. $G_{\ell}=G$ for all $\ell$.
Note however that the individual number of bath spins $M_{\ell}$
and bath couplings $g_{k}^{(\ell)}$ may still depend on $\ell$ and
$k$ as long as their means square average is the same. Also, our analytic
solution given in the next paragraph relies on this assumption, but
numerics show that our main result {[}Equation~(\ref{eq:scalingformula}){]}
remains a good approximation if the $G_{\ell}$ slightly vary and
we take $G\equiv\left\langle G_{\ell}\right\rangle .$

\paragraph*{\label{sec:Solving-the-Schr=F6dinger}Results:---}

In this paragraph, we solve the Schr\"{o}dinger equation for the model
outlined above and discuss the spin transfer functions. Firstly, let
us denote the orthonormal eigenstates of $H_{S}$ alone by \begin{equation}
H_{S}|\psi_{k}\rangle=\epsilon_{k}|\psi_{k}\rangle\quad(k=1,2\ldots,N)\end{equation}
with\begin{equation}
|\psi_{k}\rangle=\sum_{\ell=1}^{N}a_{k\ell}|x,0\rangle.\label{eq:eigen_h_c}\end{equation}
 For what follows, it is not important whether analytic expressions
for the eigensystem of $H_{S}$ can be found. Our result holds even
for models that are not analytically solvable, such as the randomly
coupled chains considered in \cite{NJP}. We now make an ansatz for
the eigenstates of the full Hamiltonian, motivated by the fact that
the states \begin{equation}
|\phi_{\ell}^{n}\rangle\equiv\frac{1}{\sqrt{2}}\left(|\ell,0\rangle+\left(-1\right)^{n}|0,\ell\rangle\right)\quad(n=1,2)\end{equation}
are eigenstates of $H_{I}^{(\ell)}$ with the corresponding eigenvalues
$\pm G$ {[}this follows directly from Eq.~(\ref{eq:hb_action})/(\ref{eq:hb_action2}){]}.
Define the vectors\begin{eqnarray}
|\Psi_{k}^{n}\rangle & \equiv & \sum_{\ell=1}^{N}a_{k\ell}|\phi_{\ell}^{n}\rangle\label{eq:ansatz}
\end{eqnarray} with $k=1,2,\ldots,N$ and $n=0,1.$ The $|\Psi_{k}^{n}\rangle$ form
an orthonormal basis in which we express the matrix elements of
the Hamiltonian. We can easily see that\begin{equation}
H_{I}|\Psi_{k}^{n}\rangle=-\left(-1\right)^{n}G|\Psi_{k}^{n}\rangle\end{equation}
and \begin{equation}
H_{S}|\Psi_{k}^{n}\rangle=\epsilon_{k}\sum_{x=1}^{N}a_{kx}|x,0\rangle=\frac{\epsilon_{k}}{2}\left(|\Psi_{k}^{0}\rangle+|\Psi_{k}^{1}\rangle\right).\end{equation}
Therefore the matrix elements of the full Hamiltonian
$H=H_{S}+H_{I}$ are given by \begin{equation}
\langle\Psi_{k'}^{n'}|H|\Psi_{k}^{n}\rangle=\delta_{kk'}\left(-\left(-1\right)^{n}G\delta_{nn'}+\frac{\epsilon_{k}}{2}\right).\end{equation}
The Hamiltonian is not diagonal in the states of
Eq.~(\ref{eq:ansatz}). But $H$ is now block diagonal consisting of
$N$ blocks of size $2$, which can be easily diagonalized
analytically. The orthonormal eigenstates of the Hamiltonian are
given by\begin{equation} |E_{k}^{n}\rangle=c_{kn}^{-1}\left\{
\left(\left(-1\right)^{n}\Delta_{k}-2G\right)|\Psi_{k}^{0}\rangle+\epsilon_{k}|\Psi_{k}^{1}\rangle\right\}
\label{eq:eigenstates}\end{equation} with the
eigenvalues\begin{equation}
E_{k}^{n}=\frac{1}{2}\left(\epsilon_{k}+\left(-1\right)^{n}\Delta_{k}\right)\end{equation}
and the normalization\begin{equation}
c_{kn}\equiv\sqrt{\left(\left(-1\right)^{n}\Delta_{k}-2G\right)^{2}+\epsilon_{k}^{2}},\end{equation}
where\begin{equation}
\Delta_{k}=\sqrt{4G^{2}+\epsilon_{k}^{2}}.\label{eq:delta}\end{equation}
Note that the ansatz of Eq.~(\ref{eq:ansatz}) that put $H$ in
block diagonal form did not depend on the details of $H_{S}$ and
$H_{I}^{(\ell)}.$ The methods presented here can
be applied to a much larger class of systems, including the
generalized spin star systems (which include an interaction within
the bath) discussed in \cite{Alexandra}.

\begin{figure}
\begin{center}\includegraphics[%
  width=0.97\columnwidth]{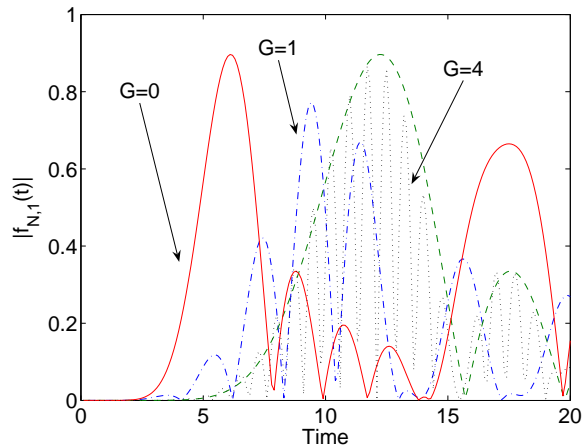}\end{center}

\caption{\label{cap:Example}The absolute value of the transport
function $f_{N,1}(t)$ of an uniform spin chain (i.e. $J_{\ell}=1$)
with length $N=10$ for three different values of the bath coupling
$G.$ The dashed line is the envelope of the limiting function for
$G\gg\epsilon_{k}/2$ given by $|f^{0}(\frac{t}{2})|.$ We can see
that Eq.~(\ref{eq:scalingformula}) becomes a good approximation
already at $G=4.$}
\end{figure}

\begin{figure}
\begin{center}\includegraphics[%
  width=0.97\columnwidth]{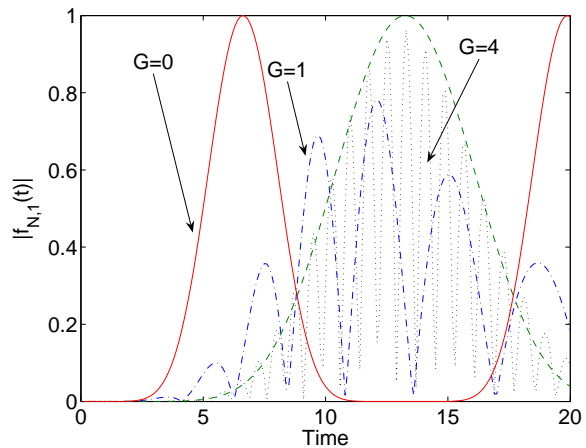}\end{center}

\caption{\label{cap:Example2}The same as Fig. \ref{cap:Example},
but now for an engineered spin chain {[}i.e.
$J_{\ell}=\sqrt{\ell(N-\ell)}${]} as in \cite{C1}. For comparison,
we have rescaled the couplings such that $\sum_{\ell}J_{\ell}$ is
the same as in the uniform coupling case.}
\end{figure}

\begin{figure}
\begin{center}\includegraphics[%
  width=0.97\columnwidth]{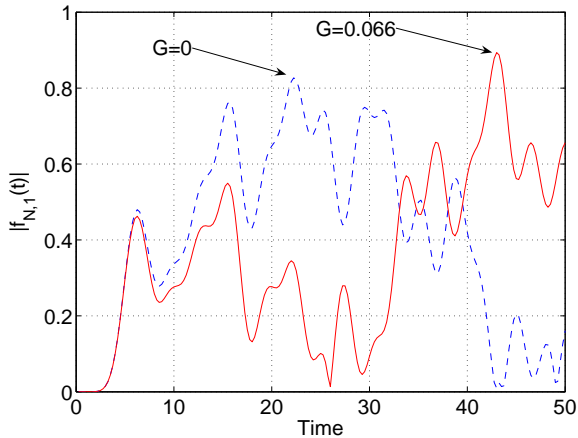}\end{center}

\caption{\label{cap:Example3}A weakly coupled bath may even improve the transfer
function for some specific choices of the $J_{\ell}.$ This plot shows
the transfer function $|f_{N,1}(t)|$ for $N=10.$ The couplings $J_{\ell}$
were found numerically. }
\end{figure}

After solving the Schr\"{o}dinger equation, let us now turn to quantum
state transfer. The relevant quantity \cite{Sougato,C1,NJP} is
given by the transfer function\begin{eqnarray*}
f_{N,1}(t) & \equiv & \langle N,0|\exp\left\{ -iHt\right\} |1,0\rangle\\
 & = & \sum_{k,n}\exp\left\{ -iE_{k}^{n}t\right\} \langle E_{k}^{n}|1,0\rangle\langle N,0|E_{k}^{n}\rangle.\end{eqnarray*}
The modulus of $f_{N,1}(t)$ is between $0$ (no transfer) and $1$
(perfect transfer) and fully determines the fidelity of state transfer.
Since\begin{eqnarray*}
\langle\ell,0|E_{k}^{n}\rangle & = & c_{kn}^{-1}\left\{ \left(\left(-1\right)^{n}\Delta_{k}-2G\right)\langle\ell,0|\Psi_{k}^{0}\rangle+\epsilon_{k}\langle\ell,0|\Psi_{k}^{1}\rangle\right\} \\
 & = & \frac{c_{kn}^{-1}}{\sqrt{2}}\left(\left(-1\right)^{n}\Delta_{k}-2G+\epsilon_{k}\right)a_{k\ell}\end{eqnarray*}
we get\begin{eqnarray}
\lefteqn{f_{N,1}(t)=}\label{eq:transfer}\\
 &  & \frac{1}{2}\sum_{k,n}e^{\frac{-it}{2}\left(\epsilon_{k}+\left(-1\right)^{n}\Delta_{k}\right)}\frac{\left(\left(-1\right)^{n}\Delta_{k}-2G+\epsilon_{k}\right)^{2}}{\left(\left(-1\right)^{n}\Delta_{k}-2G\right)^{2}+\epsilon_{k}^{2}}a_{k1}a_{kN}^{*}.\nonumber \end{eqnarray}
Eq.~(\ref{eq:transfer}) is the main result of this article, fully
determining the transfer of quantum information and entanglement
in the presence of the environments. In the limit $G\rightarrow0,$
we have $\Delta_{k}\approx\epsilon_{k}$ and $f_{N,1}(t)$
approaches the usual result \cite{Sougato,C1,NJP} without an
environment,\begin{equation}
f_{N,1}^{0}(t)\equiv\sum_{k}\exp\left\{ -it\epsilon_{k}\right\}
a_{k1}a_{kN}^{*}.\end{equation} In fact, a series expansion of
Eq.~(\ref{eq:transfer}) yields that the first modification of the
transfer function is of the order of $G^{2},$\begin{equation}
G^{2}\sum_{k}a_{k1}a_{kN}^{*}\left[\exp\left\{
-it\epsilon_{k}\right\}
\left(-\frac{1}{\epsilon_{k}^{2}}-\frac{it}{\epsilon_{k}}\right)+\frac{1}{\epsilon_{k}^{2}}\right].\end{equation}
Hence we the effect is small for very weakly coupled baths.
However, as the chains get longer, the lowest lying energy
$\epsilon_{1}$ usually approaches zero, so the changes become more
significant (scaling as $1/\epsilon_{k}$). For intermediate $G,$
we evaluated Eq.~(\ref{eq:transfer}) numerically and found that
the first peak of the transfer function generally becomes slightly
lower, and gets shifted to higher times (Figures \ref{cap:Example}
and \ref{cap:Example2}). A numeric search in the coupling space
$\left\{ J_{\ell},\ell=1,\ldots,N-1\right\} $ however also
revealed some rare examples where an environment can also slightly
improve the peak of the transfer function (Fig
\ref{cap:Example3}).
In the strong coupling regime $G\gg\epsilon_{k}/2,$ we can approximate
Eq.~(\ref{eq:delta}) by $\Delta_{k}\approx2G.$ Inserting it in
Eq.~(\ref{eq:transfer}) then becomes\begin{eqnarray}
f_{N,1}(t) & \approx & \frac{1}{2}e^{-iGt}\sum_{k}\exp\left\{ -it\epsilon_{k}\frac{1}{2}\right\} a_{k1}a_{kN}^{*}+\nonumber \\
 &  & +\frac{1}{2}e^{Gt}\sum_{k}\exp\left\{ -it\epsilon_{k}\frac{1}{2}\right\} a_{k1}a_{kN}^{*}\nonumber \\
 & = & \cos(Gt) f_{N,1}^{0}(\frac{t}{2}).\end{eqnarray}
This surprisingly simple result consists of the normal transfer function,
slowed down by a factor of $1/2,$ and modulated by a quickly oscillating
term (Figures \ref{cap:Example} and \ref{cap:Example2}). Our derivation
actually did not depend on the indexes of $f(t)$ and we get for the
transfer from the $n$th to the $m$th spin of the chain that \begin{equation}
f_{n,m}(t)\approx\cos(Gt) f_{n,m}^{0}(\frac{t}{2}).\label{eq:scalingformula}\end{equation}
 It may look surprising that the matrix $f_{n,m}$ is no longer unitary.
This is because we are considering the dynamics of the chain only,
which is an open quantum system \cite{key-36}. A heuristic interpretation
of Eq.~(\ref{eq:scalingformula}) is that the excitation oscillates
back and forth between the chain and the bath (hence the modulation),
and spends half of the time trapped in the bath (hence the slowing). If the time of the maximum
of the transfer function $|f_{n,m}^0(t)|$ for $G=0$ is a multiple of $\pi/2G$ then this maximum is also reached in the presence of the bath.

Finally, we want to stress that Eq.~(\ref{eq:scalingformula}) is \emph{universal}
for any spin Hamiltonian that conserves the number of excitations,
i.e. with $\left[H_{S},\sum_{\ell}Z_{\ell}\right]=0$.
Thus our restriction to chain-like topology and exchange couplings
for $H_{S}$ is not necessary. In fact the only difference in the
whole derivation of Eq.~(\ref{eq:scalingformula}) for a more general
Hamiltonian is that Eq.~(\ref{eq:hc_action}) is replaced by\begin{eqnarray}
H_{S}|\ell,0\rangle & = & \sum_{\ell'}h_{\ell'}|\ell',0\rangle.
\label{eq:hc_action2}\end{eqnarray}
The Hamiltonian can still be formally diagonalized in the first excitation
sector as in Eq.~(\ref{eq:eigen_h_c}), and the states of Eq.~(\ref{eq:eigenstates})
will still diagonalize the total Hamiltonian $H_{S}+H_{I}.$ Also,
rather than considering an exchange Hamiltonian for the interaction
with the bath, we could have considered a Heisenberg interaction,
but only for the special case where all bath couplings $g_{k}^{(\ell)}$ are
all the same \cite{SUBRA}. Up to some irrelevant phases, this leads to the same
results as for the exchange interaction.

\paragraph*{Conclusion:---}

We found a surprisingly simple and universal scaling law for the spin
transfer functions in the presence of spin environments. In the context
of quantum state transfer \cite{Sougato,C1,NJP} this result is double-edged:
on one hand, it shows that even for very strongly coupled baths quantum
state transfer is possible, with the same fidelity and only reasonable
slowing. On the other hand, it also shows that the fidelity as a function
of time becomes destabilized with a quickly oscillating modulation
factor. In practice, this factor will restrict the time-scale in which
one has to be able read the state from the system. This demonstrates
that even though a bath coupling need not introduce decoherence or
dissipation to the system, there are other dynamical processes such
as destabilization it may cause that can be problematic for quantum
information processing.

\paragraph*{Acknowledgments:---}

DB is funded by the UK Engineering and Physical Sciences Research
Council, Grant Nr. GR/S62796/01.

\end{document}